\documentclass[prb,twocolumn,a4paper,showpacs,superscriptaddress,floatfix]{revtex4} 

\usepackage{graphicx} 
\usepackage{amsmath}
\usepackage{bm}
\usepackage{amssymb}

\begin{document}
\draft

\title{Localized Collective Excitations in Doped Graphene in Strong Magnetic Fields}

\author{Andrea M.\ Fischer}
\affiliation{Department of Physics and Centre for Scientific Computing,
  University of Warwick, Coventry CV4 7AL, United Kingdom}
\author{Alexander B.\ Dzyubenko}
\affiliation{Department of Physics, California State University Bakersfield, Bakersfield, CA 93311, USA}
\affiliation{General Physics Institute, Russian Academy of Sciences, Moscow 119991, Russia}
\author{Rudolf A.\ R\"omer}
\affiliation{Department of Physics and Centre for Scientific Computing,
  University of Warwick, Coventry CV4 7AL, United Kingdom}

\date{$Revision: 1.112 $, compiled \today}

\begin{abstract}
We consider collective excitations in graphene with filled Landau levels (LL's) in the presence of an
external potential due to a single charged donor $D^+$ or acceptor $A^-$ impurity.
We show that localized collective modes split off the magnetoplasmon continuum and,
in addition, quasibound states are formed within the continuum. A study of the evolution
of the strengths and energies of magneto-optical transitions is performed for integer filling
factors  $\nu = 1, 2, 3, 4$ of the lowest LL.
We predict impurity absorption peaks above as well as below the cyclotron resonance.
We find that the single particle electron-hole symmetry of graphene leads to a duality between the spectra
of collective modes for the $D^+$ and $A^-$.
The duality shows up as a set of the $D^+$ and $A^-$ magneto-absorption peaks having same energies,
but active in different circular polarizations.
\end{abstract}


\maketitle

\section{Introduction}
\label{sec-introduction}

Graphene, a novel two-dimensional form of carbon,\cite{NovGMJZ04} displays exciting new physics,
distinct from that of the two-dimensional electron gas (2DEG).
A striking example in the presence of a magnetic field is the anomalous integer quantum Hall effect,
which has been observed at room temperature.\cite{NovGMJ05,ZhaTSK05}
The treatment of electronic interactions is also challenging in graphene,
due to the structure and symmetry of the dispersion relations at the two inequivalent Dirac points.
When scattering between these points is negligible, the chirality of the electrons results in a suppression of backscattering.\cite{AndNS98}
This together with graphene's reduced level of defects and impurities,
makes it highly efficient at charge transport and a promising candidate for use in nanotechnology.\cite{PonSKYH08,LinJVSF08}
However, in order for this potential to be realized, we need to better understand the precise nature of defects in graphene.
Optical measurements are a particularly useful tool for probing this.\cite{JiaHTW07,DeaCNN07,SadMPBH06,LiHJHM08}
\begin{figure*}[th]
  \centering
 \includegraphics[width=0.47\textwidth]{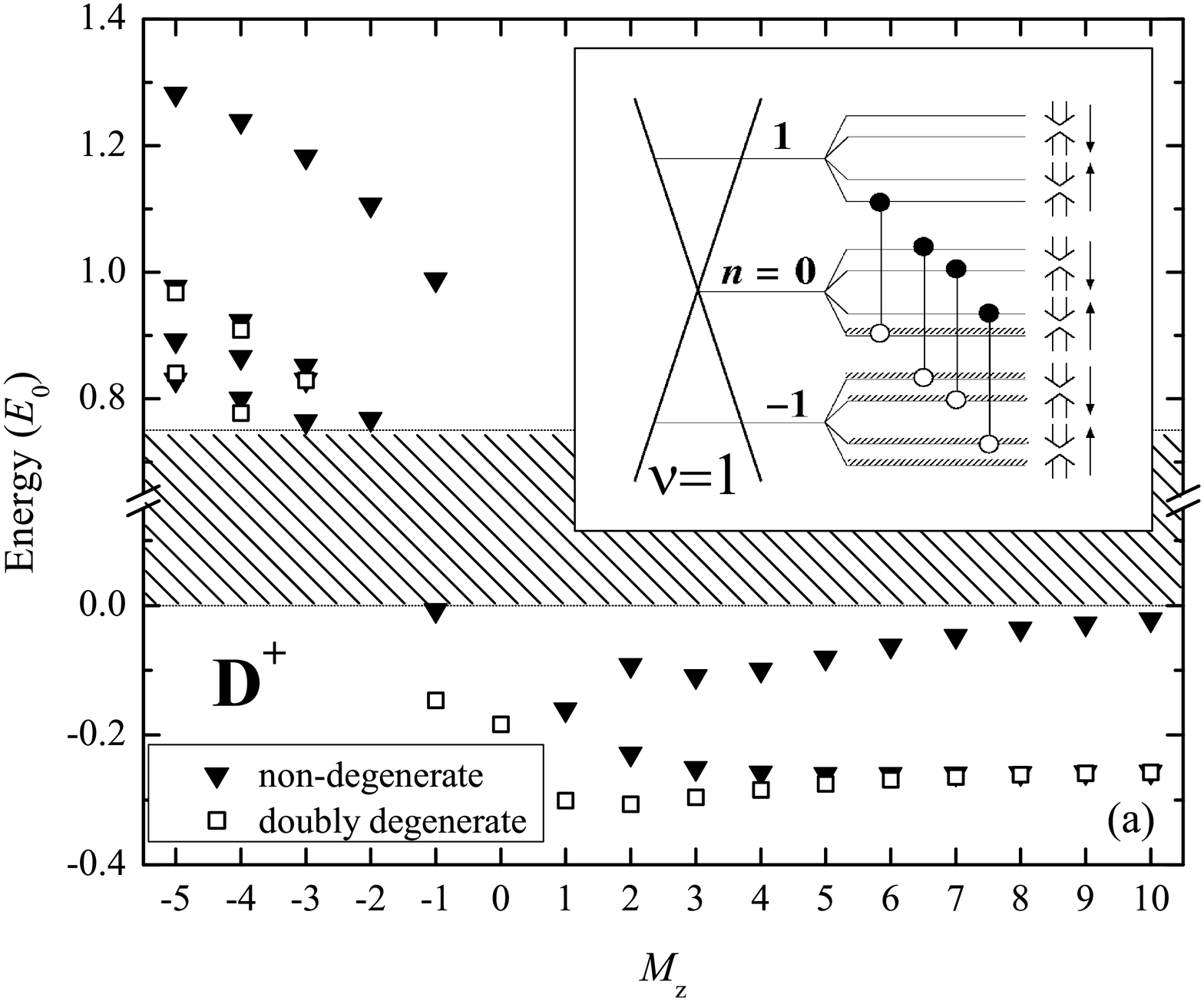}
  \includegraphics[width=0.47\textwidth]{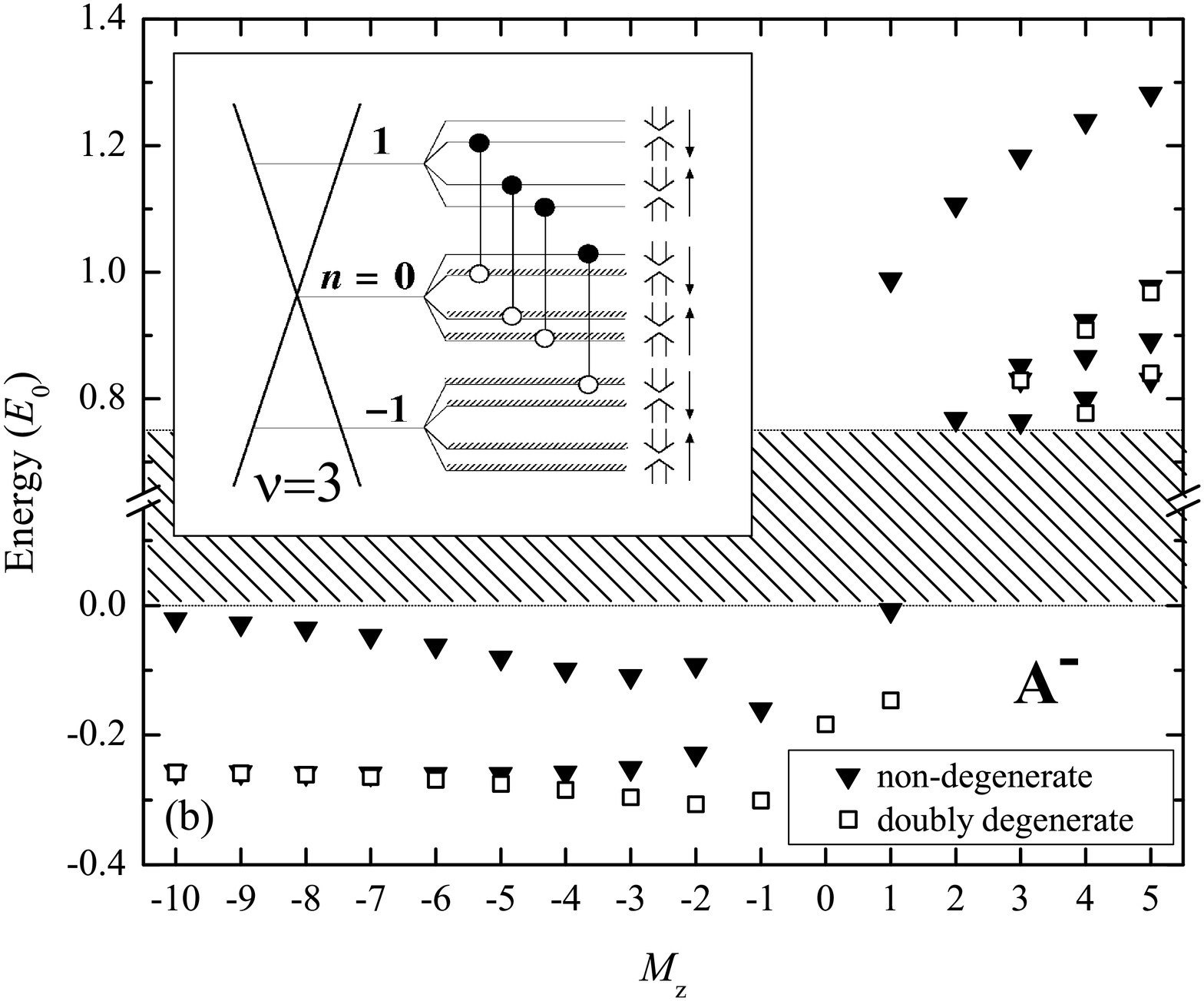}
  \caption {Magnetoplasmons bound on (a) a charged donor $D^+$ at $\nu=1$ and
(b) a charged acceptor $A^-$ at $\nu=3$.
Energies are given relative to
$\hbar\widetilde{\omega}_c$
in units of $E_0$ and $\varepsilon/\varepsilon_\mathrm{imp}=1$ (see text).
The spectra exhibit the symmetry $D^+ \leftrightarrow A^-$, $M_z \leftrightarrow -M_z$ and $\nu \leftrightarrow 4 - \nu$.
The hatched area of width $0.75 E_0$ represents the continuum of extended magnetoplasmons.
Quasibound states within the continuum are not shown.
Insets show four branches of resonantly mixed inter-LL transitions conserving spin and pseudospin.
}
\label{fig-emz}
\end{figure*}

In this paper, we study the {\em magneto-optical} response of graphene
in a strong perpendicular magnetic field
in the presence of a low density of charged impurities.
Infrared studies of Landau level (LL) transitions have reported significant departures from the bare
(non-interacting) cyclotron resonance.\cite{JiaHTW07,DeaCNN07}
Whether this can be attributed solely to interaction effects and a possible role of disorder remains unclear.
Here we develop a general formalism for studying localized collective modes of magnetoplasmon
and spin-wave types and determine their optical signatures. We treat the electron-electron
($e$-$e$) interactions beyond the conventional mean-field/RPA level and the electron-impurity interaction exactly in the high magnetic field regime.
Our results are robust for a range of impurity screening strengths.
We establish the existence of an exact symmetry for collective excitations,
which should be observable by magneto-optical spectroscopy.\cite{SadMPBH06,JiaHTW07,DeaCNN07,LiHJHM08}
This symmetry, briefly speaking, connects the magneto-optical
{\em electron}-like excitations of, e.g.\ a {\em positively} charged donor  $D^+$,
with {\em hole}-like excitations for a {\em negatively} charged acceptor  $A^-$.
This duality is a consequence of the electron-hole symmetry
between single-particle states in the lower and upper cones of graphene.\cite{CasGPNG09}
Furthermore, we establish exact optical selection rules, which demonstrate that
the ``dual'' collective excitations with total angular momenta $M_z=\pm 1$
are active in two different circular polarizations $\sigma^\pm$
and, besides having the same energies, exhibit the same oscillator strengths.
Therefore, a qualitative distinction of graphene from the conventional 2DEG,\cite{DzyL93} is that there are strong
dipole-allowed transitions in {\em both} circular polarizations
sensitive to the charge of impurity.
We show an example of this symmetry for the lowest LL $n=0$ in Fig.~\ref{fig-emz}.
Each LL in graphene consists of four sublevels, due to spin and valley (pseudospin) splitting.
We denote by $| \nu \rangle$ a many-electron ground state corresponding to the sublevel filling factor $\nu$ of a particular LL.
For sublevels $\nu=1, 2, 3$ of LL with number $n$,
the eigenstates and eigenenergies of excitations localized on the $D^+$ with $M_z$,
coincide precisely with those  with $-M_z$ formed at filling factor $\nu -4$ of the LL with number $-n$, localized on the $A^-$.
This effect represents what remains of the electron-hole symmetry after it has been broken by a charged impurity.

\section{The theoretical approach}
\label{sec-model}

We consider collective excitations in a system of electrons interacting
via a screened Coulomb potential $U_{ee}=\frac{e^2}{\varepsilon |\mathbf{r}_1-\mathbf{r}_2|}$ [\onlinecite{KalH84,IyeWFB08,BycM08}]
in the presence of an additional external field $V$.
All the results presented here are for a Coulomb impurity, $V = \pm e^2/\varepsilon_{\rm imp}|\mathbf{r}|$,\cite{imp_position}
but the method is applicable for any non-singular axially symmetric potential $V(|\mathbf{r}|)$.
The parameters $\varepsilon$ and $\varepsilon_{\rm imp}$ are the effective dielectric constants
for $e$-$e$ interaction
and electron-impurity screening,\cite{TerMKS08} respectively.
A composite index $\mathcal{N} = \{ n s \sigma \}$ is used to designate
the LL number $n$, the spin $s= \uparrow, \downarrow$, and pseudospin
$\sigma = \Uparrow, \Downarrow$ projections.
Low-energy collective excitations correspond to the promotion of one electron from one of the uppermost
filled levels $\mathcal{ N}_2$ to a higher lying empty level $\mathcal{ N}_1$ (see Fig.~\ref{fig-emz}).

Electrons in graphene follow a linear dispersion relation close to the zeroes of energy (Dirac points),
which occur at two inequivalent points in the Brillouin zone, the $\mathbf{K}$ and $\mathbf{K}'$
valleys.\cite{CasGPNG09}
We describe a perpendicular magnetic field $\mathbf{B}$
by the symmetric gauge ${\bf A} = \frac12 {\bf B} \times {\bf r}$,
consistent with the axial symmetry of $V(|\mathbf{r}|)$.
A single electron wavefunction in, e.g.\ the ${\bf K}$ valley (pseudospin $\Uparrow$),
is a four-component spinor
\begin{align}
\label{eq-Phi} 
\Phi_{n s \Uparrow m}(\mathbf{r}) = &
\langle \mathbf{r} |  c^{\dag}_{n s \Uparrow m} |0 \rangle \nonumber\\
 = & \,a_n
                     (s_n \phi_{|n|- 1 \,  m}(\mathbf{r}),
                          \phi_{|n| \,  m}(\mathbf{r}),
                                     0,
                                     0)
\chi_s \, .
\end{align}
Here, $n$ is an integer LL number, $\phi_{n  m}({\bf r})$ is a 2DEG wavefunction
with oscillator quantum number $m = 0, 1, \ldots$, $a_n=2^{\frac{1}{2}(\delta_{n,0} -1)}$,
$s_n={\rm sign}(n)$ (with $s_0=0$) and
$\chi_s$ is the spin part corresponding to two spin projections
$s = \uparrow, \downarrow$.\cite{ApaC06}
The wavefunction in the ${\bf K}'$ valley (pseudospin $\Downarrow$)
is obtained by changing the order of the spinor components.
The spinors are the eigenstates of the generalized [orbital plus isospin (sublattice)]
angular momentum projection $\hat{j}_{ze} = l_z + \frac{1}{2}\sigma_z$ [\onlinecite{DivM84}]
with half-integer eigenvalues $j_z = |n| - m - \frac{1}{2}$.
The single-electron energies are given by
$\epsilon_{\mathcal{ N}} = {\rm sign}(n) \hbar\omega_c \sqrt{|n|} + \hbar\omega_s s_z + \hbar\omega_{v} \sigma_z$,
where $\hbar\omega_c =  v_F \sqrt{2 e \hbar B/c}$
is the cyclotron energy in graphene,
$\hbar\omega_s$ is the Zeeman splitting
and $\hbar\omega_{v}$ is a possible valley splitting.\cite{ZhaJSP06}
Using the hole representation for all filled levels,
$c_{\mathcal{N} m} \rightarrow d^{\dag}_{\mathcal{N} m}$
and
$c^{\dag}_{\mathcal{N} m} \rightarrow d_{\mathcal{N} m}$
for
$\epsilon_{\mathcal{N}} \leq \epsilon_F$, we introduce operators of collective excitations as
\begin{equation}
\label{Q2} 
 \mbox{} \hspace{-5pt}
      Q^{\dag}_{\mathcal{ N}_1  \mathcal{ N}_2 M_z }  =
          \sum_{m_1 , m_2 = 0}^{\infty}
           A_{\mathcal{N}_1 \mathcal{N}_2 M_z }(m_1,m_2) \,
          c^{\dag}_{\mathcal{N}_1 m_1} d^{\dag}_{\mathcal{N}_2 m_2}
\end{equation}
with expansion coefficients satisfying the condition
$A_{\mathcal{N}_1 \mathcal{N}_2 M_z }(m_1,m_2) \sim \delta_{M_z, |n_1| - m_1 - |n_2| + m_2}$.
An exact quantum number $M_z$ is an eigenvalue of the total $\hat{J}_z = \hat{j}_{ze} + \hat{j}_{zh}$;
for neutral collective excitations $M_z$ is {\em integer} and of purely {\em orbital} nature.
It has a direct geometrical meaning\cite{DzyL93} determining the average positions of the electron and the hole relative to the impurity, i.e.
\begin{align}
\label{eq-Mz}
& \langle \mathcal{ N}_1  \mathcal{N}_2 M_z |  \mathbf{r}_h^2 - \mathbf{r}_e^2 |\mathcal{ N}_1  \mathcal{N}_2 M_z \rangle \nonumber\\
& = \left( 2\left[ M_z + 2(|n_2| - |n_1|) + 1 \right] + \delta_{n_2,0} - \delta_{n_1,0} \right) \ell_B^2\; ,	
\end{align}
where the states are defined as
$Q^{\dag}_{\mathcal{ N}_1  \mathcal{ N}_2 M_z }| \nu \rangle \equiv   |\mathcal{ N}_1  \mathcal{N}_2 M_z \rangle $.

Considering the total Hamiltonian $H = H_0 + U_{ee} + V_{\rm}$  matrix elements
$\langle \mathcal{ N}_1'  \mathcal{ N}_2' M_z | H | \mathcal{ N}_1  \mathcal{ N}_2 M_z \rangle  =
H_{ \mathcal{ N}_1  \mathcal{ N}_2 }^{ \mathcal{ N}_1'  \mathcal{ N}_2'} (M_z)$,
we find that the effective Hamiltonian is given by
\begin{eqnarray}
\nonumber
 \hat{H}_{ \mathcal{ N}_1  \mathcal{N}_2 }^{ \mathcal{N}_1' \mathcal{N}_2'}   &  = & \mbox{}
             \delta_{\mathcal{ N}_1',\mathcal{ N}_1} \delta_{\mathcal{ N}_2',\mathcal{ N}_2}
			 \sum_{m=0}^{\infty}
                                  \bigl(  \tilde{\epsilon}_{\mathcal{N}_1} + \mathcal{V}_{\mathcal{N}_1 m} \bigr)
                                   c^{\dag}_{\mathcal{N}_1 m} c_{\mathcal{N}_1 m}    \\
 \label{HamD} 
          & -  & \delta_{\mathcal{ N}_1',\mathcal{ N}_1} \delta_{\mathcal{ N}_2',\mathcal{ N}_2}
		  \sum_{m=0}^{\infty}  \bigl(   \tilde{\epsilon}_{\mathcal{N}_2} + \mathcal{V}_{\mathcal{N}_2 m} \bigr)
                                   d^{\dag}_{\mathcal{N}_2 m} d_{\mathcal{N}_2 m}  \mbox{} \;\;\;\; \mbox{} \\
\nonumber
              &   -  & \sum_{\substack{m_1 , m_2 \\ m_1' , m_2'}}
        \bar{\mathcal{W}}_{\mathcal{N}_1 m_1  \, \mathcal{N}_2 m_2}^{\mathcal{N}_1' m_1' \, \mathcal{N}_2' m_2'}
                c^{\dag}_{\mathcal{N}_1' m_1'} d^{\dag}_{\mathcal{N}_2' m_2'} d_{\mathcal{N}_2 m_2} c_{\mathcal{N}_1 m_1} \, .
\end{eqnarray}
Here $\tilde{\epsilon}_{\mathcal{N}} = \epsilon_{\mathcal{N}} + E_{SE}(\mathcal{N})$ denotes the single-particle LL energy
renormalized by $e$-$e$ exchange self-energy corrections $E_{SE}(\mathcal{N})$ [\onlinecite{IyeWFB08}].
Since Kohn's theorem is not applicable in graphene,
these corrections lead to the renormalization of the bare cyclotron energy,
$\hbar\widetilde{\omega}_c = \hbar\omega_c + \delta \hbar\omega_c$ [\onlinecite{JiaHTW07,BycM08}].
For the $n=0 \rightarrow n=1$ transition (denoted hereafter by $T_{01}$), $ \delta\hbar\omega_c$ is due only
to exchange interactions with the lower cone and
$ \delta\hbar\omega_c \simeq 0.92 \, E_0$.
Here
$E_0 = (\pi/2)^{1/2} e^2/\varepsilon l_B$ is the characteristic energy of Coulomb interactions
in strong $\mathbf{B}$, $l_B = (\hbar c/eB)^{1/2}$.

Due to the spinor form of the single-particle wavefunctions, the impurity matrix elements
in graphene are connected with those in the conventional 2DEG,\cite{DzyL93}
$V_{n m} =  \langle \phi_{n m} | V(r) | \phi_{n m} \rangle $, according to
\begin{equation}
                \label{V_imp} 
    \mathcal{V}_{\mathcal{N} m} =
  \langle \Phi_{\mathcal{N} m} | V(r)| \Phi_{\mathcal{N} m} \rangle =
                a_{n}^2 \bigl( s_{n}^2 V_{|n|-1 m} + V_{|n| m} \bigr)      \, .
\end{equation}
The two-body interaction in \eqref{HamD} consists of the direct electron-hole ($e$-$h$) attraction
and exchange $e$-$h$ repulsion, i.e.\
\begin{equation}
                \label{U_ee} 
   \bar{\mathcal{W}}_{\mathcal{N}_1 m_1  \, \mathcal{N}_2 m_2}^{\mathcal{N}_1' m_1' \, \mathcal{N}_2' m_2'}  =
                             \mathcal{W}_{\mathcal{N}_1 m_1  \, \mathcal{N}_2' m_2'}^{\mathcal{N}_1' m_1' \, \mathcal{N}_2 m_2} -
                             \mathcal{W}_{\mathcal{N}_1 m_1  \, \mathcal{N}_2' m_2'}^{\mathcal{N}_2 m_2 \, \mathcal{N}_1' m_1'} \, .
\end{equation}
In electron representation,
\begin{eqnarray}
\mathcal{W}_{\mathcal{N}_1 m_1  \, \mathcal{N}_2 m_2}^{\mathcal{N}_1' m_1' \, \mathcal{N}_2' m_2'}
&\equiv
&\langle \Phi_{\mathcal{N}_1' m_1'} \Phi_{\mathcal{N}_2' m_2'}  | U_{ee} | \Phi_{\mathcal{N}_1 m_1}  \Phi_{\mathcal{N}_2 m_2}  \rangle \nonumber\\
&=&  \delta_{s_1,s_1'} \delta_{\sigma_1,\sigma_1'} \delta_{s_2,s_2'}\delta_{\sigma_2,\sigma_2'}
\mathcal{U}_{n_1 m_1 \, n_2 m_2}^{n_1' m_1' \, n_2' m_2'} \, ,
\end{eqnarray}
note that we neglect the intervalley scattering in graphene by long-range (Coulomb) potentials. Therefore the two-particle graphene matrix elements are given by
\begin{equation}
                \label{U_2D} 
\begin{split}
   \mathcal{U}_{n_1 m_1 \, n_2 m_2}^{n_1' m_1' \, n_2' m_2'} =
          a_{n_1} a_{n_2} a_{n_1'} a_{n_2'}  &  \biggl[  U_{|n_1|  \, m_1 \,\,  |n_2| \,  m_2}^{|n_1'| \,  m_1' \,\, |n_2'| \,  m_2'}    \\
            + s_{n_1}  s_{n_1'}   U_{|n_1|-1 \, m_1 \,\, |n_2| \,  m_2}^{|n_1'|-1 \, m_1' \,\, |n_2'| \, m_2'}
            + & s_{n_2}  s_{n_2'}   U_{|n_1| \,  m_1 \,\, |n_2|-1 \, m_2}^{|n_1'|  \, m_1' \,\, |n_2'|-1 m_2'}               \\
            +   s_{n_1} s_{n_2} s_{n_1'} s_{n_2'} & U_{|n_1|-1 \, m_1 \,\, |n_2|-1 \, m_2}^{|n_1'|-1 \, m_1' \,\, |n_2'|-1 \, m_2'} \biggr] \, ,
\end{split}
\end{equation}
where
$U_{n_1 m_1 \, n_2 m_2}^{n_1' m_1' \, n_2' m_2'} = \langle  \phi_{n_1' m_1'}  \phi_{n_2' m_2'} | U_{ee}|\phi_{n_1 m_1} \phi_{n_2 m_2} \rangle$
are those used in the conventional 2DEG.
We compute the  matrix elements for lowest LL's  analytically\cite{DzyL93}
and those for arbitrary LL's numerically using Eq.~(\ref{U_2D}).

In general, an infinite number of excitations (\ref{Q2}) with the same $M_z$ are mixed by the 
$e$-$e$ interactions.
However, those with different single-particle cyclotron energies
are only weakly ($\sim E_0/\hbar\omega_c$) mixed in strong magnetic fields in graphene and can be neglected.\cite{IyeWFB08,BycM08}
Let us suppose all LL's with $n<0$
are completely filled, all LL's with $n>0$ are empty, and the four LL's with $n=0$ become successively completely filled.
We designate the corresponding filling factors as $\nu = 1, 2, 3, 4$.
For each $\nu$, there are sixteen possible inter-LL excitations involving the $n=0$ LL
as an initial or final state and which have single particle energies of magnitude $\sim \hbar\widetilde{\omega}_c$.
Here we concentrate on the four excitations for which no spin or pseudospin-flip occurs (see the insets in Fig.~\ref{fig-emz}),
as these are the only excitations which are optically dipole active.
These have the same single particle energy $\hbar\widetilde{\omega}_c$ and, therefore, are strongly (resonantly) mixed.

Let us discuss some general features of our approach. In the absence of an external potential
all magnetoplasmon states are extended and the corresponding Hamiltonian matrix is infinite, see Eq.(\ref{HamD}).
The magnetoplasmons can be labeled by a continuous quasimomentum $K$ and
their eigenenergies fill a band of width $\sim E_0$.\cite{KalH84,IyeWFB08,BycM08}
In the presence of an impurity, however, some states become localized.
Importantly, the basis states \eqref{Q2} are localized
two-particle orbitals whose distances from the impurity
increase $\sim (2m)^{1/2}\,l_B$.\cite{DzyL93}
Hence, for localized excitations the scheme is convergent so that the basis can be truncated.
We include the first $N=50$ basis states for each excitation $Q^{\dag}_{\mathcal{N}_1  \mathcal{ N}_2 M_z }$
with the total matrix size being $4N$ for four strongly mixed excitations.
The achieved accuracy in eigenergies of bound states is better than $0.1\%$.

\section{Results and discussion}
\label{sec-results}

Figure~\ref{fig-emz}(a) shows for $\nu = 1$ four low-energy branches of magnetoplasmons bound on the $D^+$
for $M_z > 0$; two of these branches are degenerate.
\begin{figure*}[tb]
  \centering
 \includegraphics[width=0.95\columnwidth]{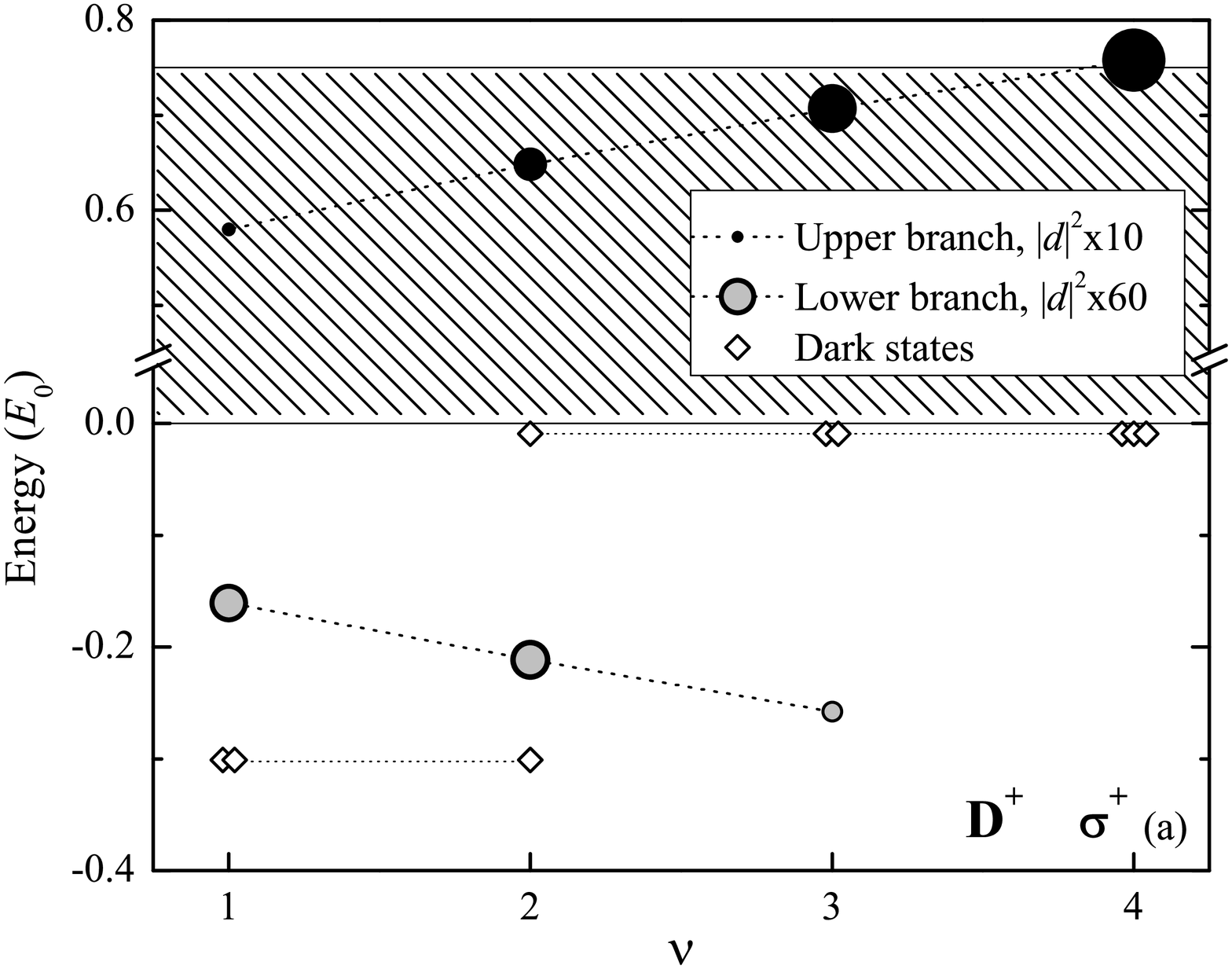}
  \includegraphics[width=0.95\columnwidth]{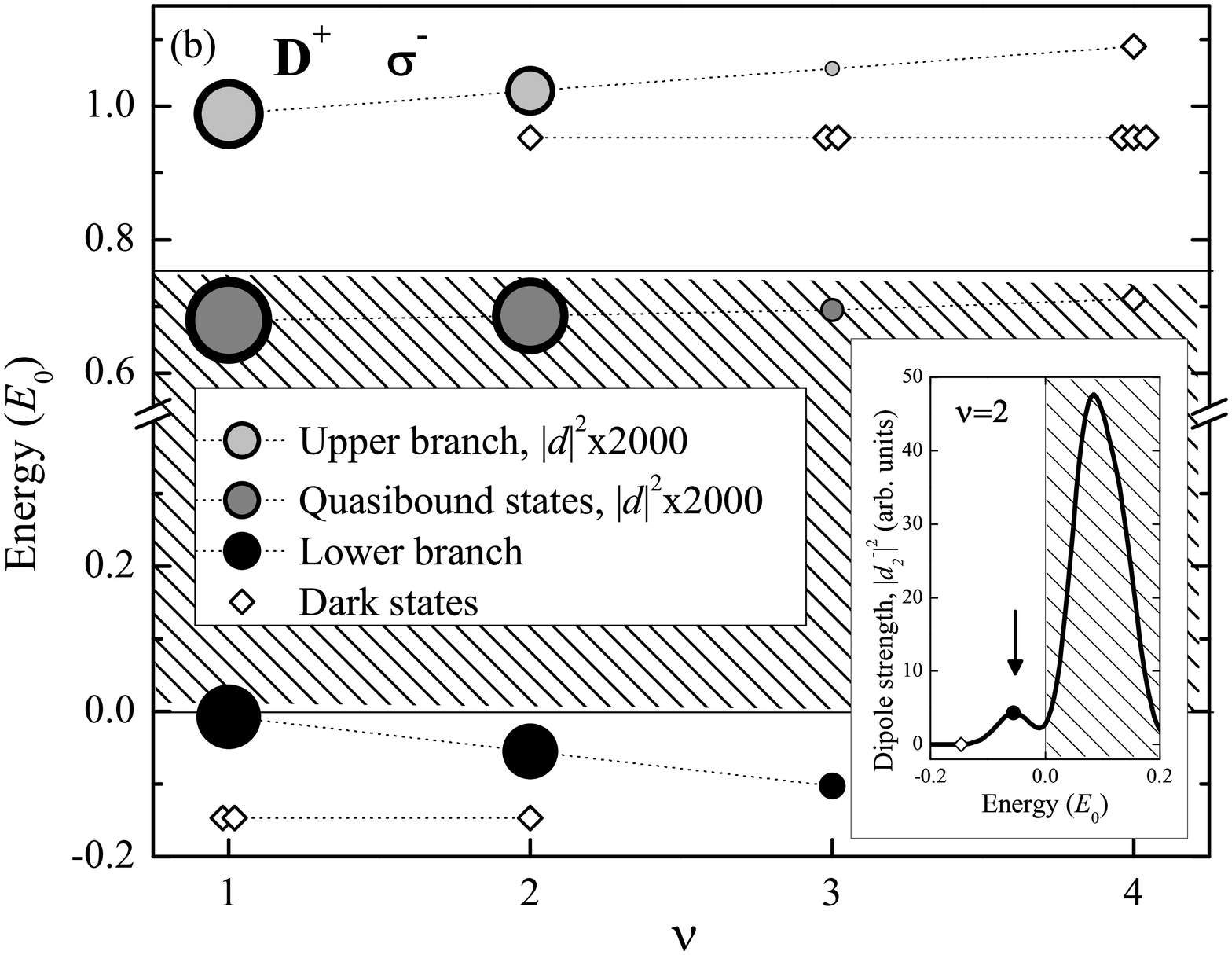}
  \caption{Evolution with filling factor $\nu$ of energies and optical strengths of magnetoplasmons bound on the $D^+$
with (a) $M_z=1$ active in the $\sigma^+$ polarization and (b) with  $M_z=-1$ active in the $\sigma^-$ polarization.
The optically active states are indicated by circles with sizes $\sim |d^{\pm}_{\nu}|^2$;
the strongest branches are shown by solid circles ($\bullet$).
The diamonds represent optically dark states.
The dotted lines are guides to the eye. Inset: Dipole strength $|d^{-}_{\nu}|^2$ vs energy for $\nu=2$.
The spectra were convoluted with a Gaussian of width $0.03E_0$.
The arrow indicates an impurity-related feature below $\hbar\widetilde{\omega}_c$
(below energy zero in the Figure).
  }
\label{fig-enu}
\end{figure*}
For large positive $M_z$, the hole is on average much farther from the impurity than the electron, see Eq.\ \eqref{eq-Mz}.
Therefore, the $e^-$-$D^+$ attraction dominates over the $h^+$-$D^+$ and $e$-$h$ interactions.
Generally, for an excitation with the electron in the $n^{\mathrm{th}}$ LL,
we find branches with asymptotic $M_z \gg 1$ energies equal to
$\mathcal{V}_{nm}$,  when counted from the corresponding renormalized cyclotron energy. 
As an example, notice the three branches approaching energy $-0.25 E_0$
and the single branch approaching
zero energy
in Fig.~\ref{fig-emz}(a).
These originate, respectively, from the three $n=-1 \rightarrow n=0$ transitions (denoted hereafter as $T_{-10}$)
and from the single $T_{01}$ transition for $\nu=1$.
Similar asymptotic behavior can be seen for other filling factors. 
The high-energy (i.e., above the band) magnetoplasmons develop for $M_z < 0$,
when the hole is on average closer to the $D^+$ than the electron.
Such unusual excited states are bound in 2D because of the confining effect of $\mathbf{B}$.\cite{DzyL93}
%
%
Fig.\ \ref{fig-emz}(b) shows the spectra for $\nu=3$ for the $A^-$ which is a ``mirror reflection'' of Fig.\ \ref{fig-emz}(a) because of the aforementioned symmetry. Generally, due to the symmetry, results for the $A^-$ at $\nu=1,2,3$ can be obtained from those for the $D^+$
by changing $M_z \rightarrow - M_z$ and $\nu \rightarrow 4 - \nu$.
For all filling factors, the spectra of bound states are qualitatively similar to those shown in Fig.\ \ref{fig-emz}.
For $\nu=4$, the states can be classified according to the total spin and pseudospin, so the states are either spin and pseudospin singlets or triplets.\cite{IyeWFB08,BycM08} Only spin and pseudospin singlets are optically active.

Let us consider the magneto-optical response in graphene.\cite{AbeF07,BycM08}
In the electric dipole approximation, the interaction of electrons with light of frequency $\omega$ and
left ($+$) and right  ($-$) circular polarizations is described by the Hamiltonian
\begin{equation}
 \delta H_{\pm} =  \frac{e v_F \mathcal{E}e^{-i\omega t}}{i\omega c}
                \left( \begin{smallmatrix}
            \sigma_{\pm}     &   0   \\
            0     &    \sigma_{\pm}
                       \end{smallmatrix} \right) \, ,
\end{equation}
where $\mathcal{ E}$ is the electric field amplitude and
$\sigma_{\pm} = \sigma_x \pm i \sigma_y$
are the Pauli matrices acting in the space of the graphene crystal sublattices.
The exact optical selection rules for the collective excitations that follow are:
only those with no spin or pseudospin flips and with $M_z=\pm 1$ and $|n_1| - |n_2|= \pm 1$
are optically active in the two circular polarizations $\sigma^{\pm}$.
We quantify the rate of microwave absorption in the $\sigma^{\pm}$ polarization
by calculating the dipole transition matrix elements
$|d^{\pm}_{\nu}|^2= | \langle M_z= \pm 1 |  \delta H_{\pm} | \nu \rangle |^2$
to final states of magnetoplasmons obtained by numerical diagonalization.

Figure~\ref{fig-enu} shows optical properties of states bound on $D^+$ for the $\sigma^+$ and $\sigma^-$ polarizations.
The results for the $A^-$ at $\nu=1,2,3$ can be obtained from those reported here with the change $\nu \leftrightarrow 4-\nu$ and
$\sigma^+ \leftrightarrow \sigma^-$.
Two types of localized states can be optically observed:
(i) truly bound states, which are split off the continuum and have normalizable wavefunctions,
(ii) quasibound states within the continuum, which have high probability amplitudes on the impurity and long-range oscillating tails.
The latter may exhibit {\em asymmetric} Fano-type optical signatures.\cite{Fan61}

For both polarizations, the upper branch originates mostly from the $T_{01}$ transitions with some small
(zero at $\nu=4$) admixture of the $T_{-10}$.
With increasing $\nu$, the number of $T_{01}$ transitions increases,
which leads to the enhanced contribution of the repulsive $e$-$h$ exchange interactions.
This explains the {\em blue shift} of the upper branch to higher energies with increasing $\nu$.
Also, its optical strength $|d^{+}_{\nu}|^2$  increases (Fig.~\ref{fig-enu}a)
while $|d^{-}_{\nu}|^2$ decreases (Fig.~\ref{fig-enu}b). This is explained by the fact that
in the $\sigma^+$ ($\sigma^-$) polarization only the electron-like $T_{01}$ (hole-like $T_{-10}$) transitions are optically active.

\begin{figure}[tb]
  \centerline{
 \includegraphics[width=0.95\columnwidth]{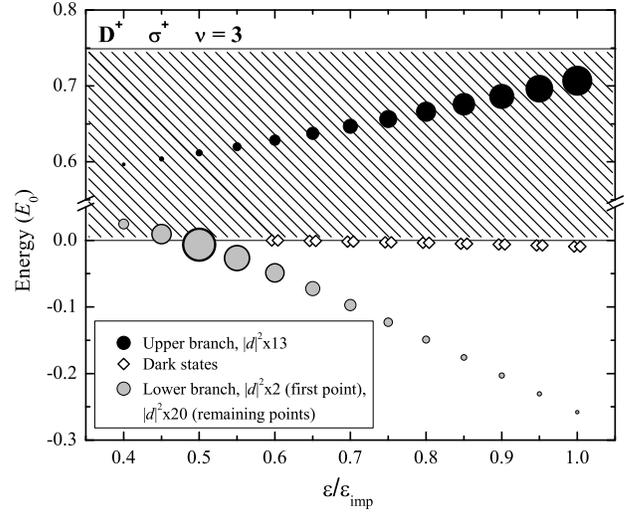}  
  }
  \caption{Dependence of energies and optical strengths of magnetoplasmons
  on ratio of effective dielectric constants $\varepsilon/\varepsilon_{\rm imp}$
  for the $D^+$ in $\sigma^+$ polarization for $\nu=3$. Symbols are as in Fig.~\ref{fig-enu}.}
\label{fig-emz-don}
\end{figure}

Screening was shown to be relevant in graphene,\cite{TerMKS08}
although the situation is not yet fully understood, particularly for a strong magnetic field.
Coulomb impurities with charge $Z=1$ belong to the subcritical regime and hence screening effects
due to the substrate and the electron system in graphene can be modelled
via an effective charge $Z_{\rm eff} < Z$,\cite{TerMKS08}
or by an effective dielectric constant $\varepsilon_{\rm imp}$.
However the value relative to the $e$-$e$ screening constant,
$\varepsilon/\varepsilon_{\rm imp}$ is unknown for graphene.\cite{IyeWFB08,BycM08}
We use $\varepsilon/\varepsilon_{\rm imp}=1$ in Figs.~\ref{fig-emz}~and~\ref{fig-enu} and show in Fig.~\ref{fig-emz-don}
how the energies of the excitations for, e.g.\ the $\nu=3$ case of Fig.~\ref{fig-enu}(a),
are modified by $\varepsilon/\varepsilon_{\rm imp}\leq 1$. As this ratio increases,
the branches of optically active bound states
are pushed away from the band of extended magnetoplasmons.

\section{Conclusions}
\label{sec-conclusions}

In conclusion, we established the spectra and the symmetries of collective excitations
bound on charged impurities in graphene in magnetic fields.
Our single-impurity theory is applicable to samples with finite impurity density $n_{\rm imp} < 1/\pi l_{B}^2$,
i.e.\ when the mean separation between impurities exceeds the size of bound magnetoplasmons.
The intensity of impurity peaks will then be $I_{\nu}^{\pm} \sim n_{\rm imp} |d_{\nu}^{\pm}|^2$.
Recent progress in fabrication of large graphene films\cite{Gei09} with sizes exceeding the wavelength
$\lambda \approx 2\pi c/ \sqrt{\varepsilon } \, \tilde{\omega}_c$
($> 60-100$\,$\mu m$)
opens the way for detailed studies of the effects predicted here.
Polarization resolved magneto-optical spectroscopy and
cyclotron resonance detection using the photoconductive response may be very effective experimental probes.
Our results demonstrate the breaking of particle-hole symmetry in a sample with predominantly positive or negative impurities,
which may partly explain the observed asymmetry in LL transitions.\cite{DeaCNN07}
The developed method can be extended for defects with short-range potentials, results will be published elsewhere.\cite{FisDR09c}


\acknowledgements
We thank D.\;N.\ Basov, V.\;I.\ Fal'ko, and M.\;M.\ Fogler for useful discussions.
We acknowledge funding by EPSRC and the Warwick North American Travel fund (AMF).
AMF is grateful for hospitality at CSU Bakersfield.
ABD acknowledges Cottrell Research Corporation and the Scholarship of KITP, UC Santa Barbara.



\begin{thebibliography}{10}
\bibitem{imp_position} The characteristic size of the localized excitations is of the order of the magnetic length, $l_B$.
Since $l_B>>a$ ($a$, the lattice spacing), the exact position of the Coulomb impurity in a primitive cell is irrelevant.
\end{thebibliography}

\end{document}